# Oscillatory spin-orbit torque switching induced by field-like torques

Jong Min Lee[1,2][†], Jae Hyun Kwon[1][†], Rajagopalan Ramaswamy[1,2], Jung Bum Yoon[1,3], Jaesung Son[1], Xuepeng Qiu[4], Rahul Mishra[1,2], Shalabh Srivastava[1], Kaiming Cai[1], and Hyunsoo Yang[1,2]*

[1]Department of Electrical Engineering and Computer Engineering, National University of Singapore, 117576, Singapore

[2]NUSNNI, National University of Singapore, 117411, Singapore

[3]Center for Nanometrology, Korea Research Institute of Standards and Science, Daejeon 34113, Republic of Korea

[4]Shanghai Key Laboratory of Special Artificial Microstructure Materials & School of Physics Science and Engineering, Tongji University, Shanghai 200092, China

[†]These authors contributed equally to this work.

Correspondence and requests for materials should be addressed to H.Y. (e-mail: eleyang@nus.edu.sg).

**Deterministic magnetization switching using spin-orbit torque (SOT) has recently emerged as an efficient means to electrically control the magnetic state of ultrathin magnets. The SOT switching still lacks in oscillatory switching characteristics over time, therefore, it is limited to bipolar operation where a change in polarity of the applied current or field is required for bistable switching. The coherent rotation based oscillatory switching schemes cannot be applied to SOT because the SOT switching occurs through expansion of magnetic domains. Here, we experimentally achieve oscillatory switching in incoherent SOT process by controlling domain wall dynamics. We find that a large field-like component can dynamically influence the domain wall chirality which determines the direction of SOT switching. Consequently, under nanosecond current pulses, the magnetization switches alternatively between the two stable states. By utilizing this**

**oscillatory switching behavior we demonstrate a unipolar deterministic SOT switching scheme by controlling the current pulse duration.**

**Introduction**

Magnetism plays a key role in modern data storage and advanced spintronics devices as non-volatile information can be encoded in the magnetization state of a nanoscale magnet. Over the last two decades, electrical methods to control the magnetization state have received immense research attention to meet the demand for the reduction in the size and energy consumption of magnetic storage cells and devices. As a result, remarkable developments have been made in switching the magnetization electrically using spin-transfer torque (STT)[1-5], electric field[6-10], and the recently discovered spin-orbit torque (SOT)[11-15]. The operation principle of majority of these electrical techniques is based on the control of the polarity of the external force, such as an electric current or magnetic field, to achieve switching between two magnetic states (bipolar switching techniques).

On the other hand, magnetization switching techniques with a fixed polarity of the external force (unipolar operation) are receiving immense attention because of their scientific interest as well as their potential to significantly increase the scalability of spintronics devices by replacing transistors with diodes[10, 16]. It is possible to achieve unipolar magnetization switching by exploiting the temporal evolution of magnetization switching, or, in other words, the magnetization dynamics. For example, in the context of STT or electric field induced magnetization switching, unipolar operation was previously demonstrated[7, 9, 16-19] by driving the magnetization into coherent precessional motion between the two stable states and then precisely controlling the time duration for which the external force of fixed polarity was applied. As the time varying magnetization trajectory oscillates between the two potential minima (stable states), switching was accomplished by removing the external force at the appropriate time to release the magnetization at the desired final state. This oscillatory behavior was also theoretically predicted in the SOT driven switching[20, 21]. However, it must be emphasized that the above described oscillatory switching scheme for unipolar operation requires strong coherency of the magnetic moments or, in other words, the magnetization should be kept uniformly aligned throughout the rotational switching process. The coherency of the magnetization is very sensitive to thermal agitations[22] and also relatively weak[23-25] in magnetic structures with perpendicular magnetic anisotropy (PMA) which are required for high density applications.

Here, we report our experimental discovery of an alternative method to achieve the oscillatory switching behavior in the scenario of incoherent magnetization switching in PMA structures driven by SOT. The SOT is an electric current induced phenomenon that utilizes spin currents generated by spin orbit interactions to efficiently manipulate and switch the magnetization of an ultrathin magnet[11-15]. While its microscopic origin is still controversial[26-29], SOT is known to be composed of two components, namely, the damping-like torque (DLT), $\tau_{DLT} \sim \hat{m} \times (\hat{m} \times \hat{y})$ and the field-like torque (FLT), $\tau_{FLT} \sim \hat{m} \times \hat{y}$. Here, $\hat{m}$ and $\hat{y}$ indicate the direction of the magnetization of the ultrathin magnet and the spin polarization of the incoming spin current, respectively. When substantial magnetization is orthogonal to $\hat{y}$, the DLT and FLT can be considered as equivalent field with $\hat{m} \times \hat{y}$ symmetry ($H_{DLT}$) and $\hat{y}$ symmetry ($H_{FLT}$), respectively[30].

Unless the lateral dimensions of the ultrathin magnet are extremely small (< 40 nm), the SOT induced magnetization switching in PMA structures is an incoherent process[31]. In the incoherent regime, the switching happens by depinning of a reversed magnetic domain followed by its expansion[24, 30-35]. Due to the torque symmetries, the general consensus till now is that the DLT is responsible to drive the domain expansion in SOT switching[30] and the role of FLT in deterministic switching is usually neglected and not well understood[25, 30]. On the contrary, our studies reveal that in PMA structures with large FLT, the SOT driven incoherent magnetization dynamics and the deterministic switching are greatly influenced by FLT.

**Results**

**Nanosecond pulse current induced SOT switching**

We explore the SOT driven magnetization switching dynamics under the application of nanosecond current pulses in Ta/CoFeB/MgO structures, whose FLT is large and is of opposite sign to that of DLT[27, 36] (FLT/DLT = –3.2, see Methods for our sign conventions). As shown in Fig. 1a, a perpendicularly magnetized circular dot with 1000 nm diameter ($d$) was patterned on top of Ta channel. The electric current pulses are applied along the +$x$-direction and an in-plane assist field ($H$) is applied in the $xy$-plane where its in-plane angle ($\theta_H$) is defined with respect to the +$x$-axis. The applied $H$ is along the –$x$-direction ($\theta_H = 180°$) unless otherwise specified. The details of device preparation and measurement are described in the Methods.

In order to study the SOT switching dynamics, we have measured the probability of magnetization switching by applying current pulses with the initial state of the magnetization as 'up' (+$z$-direction). Figure 1b shows the two dimensional diagram of the measured switching probability ($P_{sw}$) as a function of current density ($J$) and pulse duration ($t$) at a fixed $H$ = 1191 Oe. We have also measured the $P_{sw}$ vs. $t$ for different $H$ while keeping a constant value of $J$ (79.4 × 10$^6$ A cm$^{-2}$) as shown in Fig. 1c. Under the application of the current pulses, a clear 'up' to 'down' SOT switching is observed as indicated by the transition of $P_{sw}$ from 0 % to 100 %. This first switching boundary (where $P_{sw}$ = 50 %) between the initial state and forward switching is monotonic with respect to $J$, $t$ and $H$, suggesting that the forward switching is more likely to occur with a larger $J$, a longer $t$ or a larger $H$, which is expected from torque driven SOT magnetization switching dynamics as observed before[24, 31]. Moreover, the 'up' to 'down' switching direction is also consistent with the previous experiments and also with that of DC-current induced SOT switching in our devices (see Supplementary Note 1 and 2). By performing a linear fit of the critical switching current density (at the first switching boundary) with corresponding values of 1/$t$, we estimate the intrinsic critical switching current density ($J_{c0}$) in our devices as $43.2 \times 10^6$ A cm$^{-2}$ (see Supplementary Note 3). This value of $J_{c0}$ is significantly smaller than the calculated value of $148 \times 10^6$ A cm$^{-2}$ from the macrospin-like coherent switching model (see Methods for details), suggesting that the switching in our device occurs via expansion of reversed domain[24, 30-35] rather than coherent magnetization rotation, which is also expected from the size of the studied structure.

**Oscillatory switching behavior induced by FLT**

Beyond the first switching boundary, the $P_{sw}$ is expected to remain at 100 % and does not change, since the existing theories and experimental results indicate that the DLT driven incoherent SOT switching is a deterministic process[24, 30, 31]. On the contrary, as seen in Figs. 1b and 1c, if we apply a pulse with a longer $t$, a backward switching boundary appears where the magnetization flips back from 'down' to its initial 'up' state. This unexpected backward switching observed in our devices, for a wide range of $J$, $t$, and $H$, is also a spin torque driven process, since the backward switching boundary also shows a monotonic behavior with $J$, $t$ and $H$. On applying a longer $t$ beyond the backward switching, the magnetization undergoes forward switching again (from 'up' to 'down' state) resulting in an oscillatory behavior of $P_{sw}$. We note that the backward switching and the oscillatory behavior are observed regardless of the initial magnetization states ('up' and 'down', see Supplementary Note 4).

The occurrence of oscillatory $P_{sw}$ is surprising because the SOT switching in our devices proceeds by domain expansion unlike the previous reports where the switching takes by coherent magnetization rotation[7, 9, 18-21]. Furthermore, the signature of incoherent switching in our devices can be also observed in the oscillatory period of $P_{sw}$. In the case of coherent switching (see Supplementary Note 5), the oscillatory period of $P_{sw}$ is quite symmetric as it arises from the precessional motion with a constant frequency (~ Larmor frequency). On the other hand, the observed periods in our study are distinctly asymmetric as the observed period for the backward switching is much longer than that for the first forward switching. For instance, the periods of the first forward switching and backward switching are ~2.7 ns and ~7.5 ns, respectively, for an applied $J$ of 79.4 × 10$^6$ A cm$^{-2}$ and $H$ of 1191 Oe, which are indicated by dashed arrows in Fig. 1b.

In order to obtain more insights on the backward switching, we have measured $P_{sw}$ for different $\theta_H$ as shown in Fig. 1d. Interestingly, the observed 'down' to 'up' backward switching exhibits significant asymmetric behavior with respect to $\theta_H$, compared to the 'up' to 'down' forward switching. The backward switching is suppressed or enhanced, as the $H$ is tilted towards ($\theta_H < 180°$) or away from ($\theta_H > 180°$) the +y-direction, respectively. This asymmetric behavior implies that an equivalent field with y-symmetry gives rise to the observed backward switching, and this y-symmetry coincides with the direction of $H_{FLT}$. The harmonic Hall voltage measurements in the Ta/CoFeB/MgO structure have shown that a large $H_{FLT}$ exists in the −y-direction when a positive current (along the +x-direction) is applied[27, 36]. The observed backward switching in Fig. 1d is suppressed when the effective $H_{FLT}$ is reduced by applying an external transverse field along the $+y$-direction ($\theta_H < 180°$) opposite to the SOT induced $H_{FLT}$ (along the $-y$-direction). Therefore, the contributions of FLT play a dominant role in breaking the determinism in SOT switching dynamics and thus should not be neglected. Complete suppression of the backward switching under titled $H$ toward the $+y$-direction is observed in another device (see Supplementary Note 6). Furthermore, the backward switching or the oscillatory switching behavior is not observed in the Pt layer based device which exhibits a small FLT/DLT ratio of $-0.5$ (see Supplementary Note 7).

We have then estimated and compared the domain wall (DW) velocity during the first forward and backward switching processes. The mean DW velocity ($V_{DW}$) during the forward switching is estimated using the relation, $V_{DW,fwd} = d/(2t_{c,fwd})$ with an assumption of SOT switching occurs by reverse domain nucleation at one corner followed by its expansion across the PMA dot [30, 32, 33, 35]. Here, $t_{c,fwd}$ represents the time corresponding to $P_{sw} = 50\%$ during the first forward switching. $V_{DW,fwd}$ is estimated only in the relatively large $J$ regime ($J > J_{c0}$), where the spin-torque is dominant over the thermal activation[24, 37]. As shown in Fig. 2a, the estimated $V_{DW,fwd}$ shows a proportional increase with an increase in $J$, and we obtain a $V_{DW,fwd}$ of 504 m s$^{-1}$ for $J = 10^8$ A cm$^{-2}$ and $H = 1191$ Oe, which is in agreement with the reported value under a large longitudinal field[38]. Figure 2b shows that the $V_{DW}$ increases with increase in the magnitude of $H$ which can be understood as follows. As $H$ increases, the magnetization at the center of the domain wall ($\mathbf{M_{DW}}$) is better aligned towards the $H$ direction ($-x$ direction). Subsequently, the out-of-plane $H_{DLT}$ ($\propto \mathbf{M_{DW}} \times \hat{y} \propto x$-component of $\mathbf{M_{DW}}$) exerted on the DW also increases leading to a larger $V_{DW}$[14, 30]. Figure 2c shows the $V_{DW,fwd}$ as a function of the transverse component ($y$ component) of the applied $H$ (top axis). The corresponding FLT$_{eff}$/DLT ratio is indicated in the bottom axis, which is defined from following relation: $\left(H_{FLT}(J) - H\cos\theta_H\right)/H_{DLT}(J)$. Here, $H_{DLT}(J)$ and $H_{FLT}(J)$ are the corresponding SOT fields at a given current density which are measured from the harmonic technique (see Supplementary Note 8). The asymmetric behavior of $V_{DW,fwd}$ with respect to the transverse component of $H$ arises due to **M**$_{DW}$ being pulled away (into) the Néel wall configuration resulting in decrease (increase) of the $H_{DLT}$ experienced by the DW[14].

Similarly, we have determined the $V_{DW}$ during the observed backward switching using the relation, $V_{DW,bck} = d/2(t_{c,bck} - t_{c,fwd})$, as the backward switching follows the first forward switching in time. The $t_{c,bck}$ represents the time corresponding to $P_{sw} = 50\%$ during the backward switching. Interestingly, the estimated $V_{DW,bck}$ also shows monotonic increase with respect to $J$ and $H$ (Figs. 2a and 2b) and an asymmetric behavior as a function of $\theta_H$ (Fig. 2c), implying that the backward switching also arises from the spin torque driven domain expansion similar to the case of the first forward switching but in an opposite manner. However, $V_{DW,bck}$ is smaller than $V_{DW,fwd}$ because the domain expansion in the backward switching is energetically unfavorable as discussed later.

**One-dimensional micromagnetics simulations of domain walls**

In order to understand the experimental observations and elucidate the role of FLT in the oscillatory $P_{SW}$, we have performed one-dimensional (1D) micromagnetics simulations of the SOT switching driven by domain expansion (see Methods for details). The top panel of Fig. 2d shows the SOT induced temporal evolution of averaged out-of-plane magnetization ($m_z$) as a function of the FLT/DLT ratio, where the value of DLT is kept constant. At the start of the simulation (0 ns), a reversed 'down' domain is introduced at one edge of the structure. This reversed domain is then expanded by SOT as the simulation proceeds. In the case where there is no FLT, the SOT successfully switches the magnetization to 'down' ($m_z = -1$) state. However, when a large FLT is considered (FLT/DLT = $-4.8$), the 1D model also reproduces the backward switching behavior as the $m_z$ returns back to a positive value after the forward switching. This backward switching behavior is gradually suppressed with decreasing the magnitude of the FLT/DLT ratio, which is consistent with the experimental observation. Figures 2e-g show the calculated $V_{DW}$ during the forward and backward switching as a function of $J$, $H$, and FLT/DLT ratio, respectively. The calculated and experimentally determined $V_{DW}$ also show good qualitative agreement as the monotonic behavior with respect to $H$ and $J$, asymmetric behavior with respect to the FLT/DLT ratio and the slower velocity during backward switching are reproduced.

The bottom panel of Fig. 2d shows the temporal evolutions of azimuthal angle of DW ($\theta_{DW}$), which is the angle between $\mathbf{M_{DW}}$ and $+x$-direction. The evolution of $\theta_{DW}$ for the different ratios of FLT/DLT sheds light on the key role of FLT on the domain expansion in the opposite direction and the resultant backward switching. At the start of simulation, due to the applied $H$, the $x$-component of $\mathbf{M_{DW}}$ is along the $-x$-direction and thus $\theta_{DW} = 180°$. Under the application of SOT, the reversed domain expands and $\theta_{DW}$ gradually decreases to 90° as $\mathbf{M_{DW}}$ damps towards the spin polarization direction[39]. For the case without FLT (FLT/DLT = 0), the DW annihilates as it expands to the structure edge ($m_z = -1$) which results in $\mathbf{M_{DW}}$ and thus $\theta_{DW}$ not being well defined. However, when a sizeable FLT of opposite sign to DLT is considered, $\theta_{DW}$ exhibits an oscillatory behavior over time, which indicates that the DW does not immediately annihilate after it reaches the structure edge. Further, it is observed that the time for which |$\theta_{DW}$| is stable below 90° increases with increasing the magnitude of FLT and as we explain in the following

paragraph, whenever the value of |$\theta_{DW}$| < 90° ($\mathbf{M_{DW}}$ in the +x-direction), the SOT drives the backward switching. This result indicates that the FLT facilitates backward switching by stabilizing |$\theta_{DW}$| < 90°.

The physics behind the FLT induced oscillatory behavior of $\theta_{DW}$ and the resultant backward switching is illustrated in Fig. 3 using the DW configuration, $\mathbf{M_{DW}}$ orientation, and torques acting on $\mathbf{M_{DW}}$ at different times. Time 1 corresponds to the case for 'up' to 'down' forward switching process when the x-component of $\mathbf{M_{DW}}$ is stabilized along −x-direction ($\mathbf{M_{DW}} \cdot \hat{\mathbf{x}} < 0$). Consequently, the DW experiences an out-of-plane $H_{DLT}$ in the −z-direction ($\mathbf{M_{DW}} \times \hat{\mathbf{y}} < 0$) and the 'down' domain expands to advance the forward switching process. Time 2 corresponds to the case when the propagating DW reaches the structure edge and annihilates. However, this annihilation process is followed by a nucleation of a DW with an inverted chirality ($\mathbf{M_{DW}} \cdot \hat{\mathbf{x}} > 0$) which can be understood as a reflection of the DW on the structure edge[40, 41]. This DW with an inverted chirality is not energetically favorable and follows damped motion over time to revert back its chirality due to the applied $H$ along the −x-direction. However, a sufficiently large $H_{FLT}$ in the −y-direction can give dynamic stability to the DW with inverted chirality with a lifetime of several nanoseconds. As this metastable DW's center is along the +x-direction, it experiences a $H_{DLT}$ in the +z-direction ($\mathbf{M_{DW}} \times \hat{\mathbf{y}} > 0$), therefore the 'up' domain expands, as shown in time 3, which results in the backward switching. Over time, the metastable DW recovers back its chirality with its $\mathbf{M_{DW}}$ again pointing back to the −x-direction which proceeds to switch the magnetization again in the forward direction and the whole cycle repeats giving rise to the oscillatory behavior in $P_{sw}$. The velocities of the two switching processes are different since the inverted DW configuration during the backward switching is in an energetically unfavorable state as the applied external $H$ is against $\mathbf{M_{DW}}$. Furthermore, the attained metastability of the reversed DW decreases over time and eventually only the forward switching will prevail. As a result, the backward switching can be observed only in the nanosecond time scale.

Although the DW chirality and the resulting domain expansion are discussed in terms of $H$ and $H_{FLT}$, it is known that the other effects such as Dzyaloshinskii-Moriya interaction (DMI) may also influence the DW chirality. It is reported that the DMI plays a central role in the case of SOT driven DW displacement, as it determines the DW chirality and the sign of out-of-plane $H_{DLT}$ experienced by DW[14, 38, 42, 43]. However, in the case of SOT driven switching, the DW chirality is reported to be governed largely by the applied $H$, as it overcomes the DMI effective field[30, 43]. In the studied structure, the DMI effective field is estimated as 103 Oe (see Methods for details) and is quite smaller than the $H$ (550 ~ 1200 Oe) and $H_{FLT}$ (1257 Oe per $79.4 \times 10^6$ A cm$^{-2}$, see Supplementary Note 8) applied in the experiments and simulations. Therefore, we believe that the DW chirality during the forward and backward switching is dominantly governed by $H$ and $H_{FLT}$.

**Unipolar SOT switching**

Finally, utilizing the observed oscillatory characteristics, we show a deterministic unipolar SOT switching scheme which reversibly controls the magnetization configuration under a constant $J$ and $H$ of a fixed polarity and changing $t$ only. This is demonstrated using a series of current pulses with alternating lengths of 2.5 ns and 7.5 ns with a fixed current density of $79.4 \times 10^6$ A cm$^{-2}$ under a constant $H$ of 1067 Oe. After each pulse injection, the magnetization state is monitored using the anomalous Hall resistance ($R_{AHE}$) measurement. As shown in Fig. 4, the deterministic SOT switching consistently occurs by the unipolar current pulses. The initial state of the magnetization is pointing 'up' and the pulse of 2.5 ns always switches the magnetization to 'down', while the magnetization is always brought to 'up' state with the pulse of 7.5 ns.

**Discussion**

The role of FLT has not been paid much attention in the majority of SOT switching experiments and thus, the SOT switching and domain wall dynamics have been mainly discussed using the DLT alone. The FLT was claimed to induce a partial decrease in the SOT switching probability (decreased to ~60% after achieving full 100% switching)[44]. However, another work reported a similar backward SOT switching that was attributed to a small tilt of in-plane assist field along the out-of-field direction[45]. With these two different interpretations, the role of FLT in SOT dynamics and the underlying physics of the backward SOT switching still remain vague. In this regard, our experiments bring to light the crucial role of FLT in breaking the determinism in SOT driven incoherent switching dynamics which results in oscillatory magnetization switching characteristics with respect to the current pulse duration. We make use of this observed oscillatory behavior to demonstrate a unipolar deterministic SOT switching scheme which operates by controlling the duration of the current pulses, while keeping the magnitudes and polarities of the current and the assist-field constant. Our study provides the missing piece in the physics of SOT switching dynamics and offers novel strategies for magnetization switching with unipolar operation.

**Methods**

**Sample preparation and measurements** The film structure of Ta (6 nm)/$Co_{40}Fe_{40}B_{20}$ (0.9 nm)/MgO (2 nm)/$SiO_2$ (3 nm) on a Si/$SiO_2$ substrate is prepared by magnetron sputtering (base pressure $< 1 \times 10^{-8}$ Torr) and annealed at 200 °C for 30 minutes to improve PMA. The structure is subsequently patterned into a CoFeB circular dot with a 1000 nm diameter on top of the Ta Hall cross using electron beam lithography and Ar ion etching. The negative tone electron-beam resist of ma-N 2403 with fine resolution of 5 nm was used for patterning the Hall cross and the circular dot. The electrodes were prepared using positive tone electron-beam resist of PMMA 950 and deposition of Ta (5 nm)/Cu (100 nm). The Ta channel surface is cleaned using Ar ion etching prior to electrode deposition for Ohmic contact. The thickness of the Ta channel after fabrication is ~3.5 nm, estimated from channel resistance measurements. The films have a saturation magnetization of $M_s = 670$ emu cm$^{-3}$ and an effective anisotropy field $H_{k,eff} = 3000$ Oe measured using vibrating sample magnetometer.

DC- and nanosecond current pulses are applied in the Ta channel through a bias-tee and the perpendicular magnetization state is measured from the anomalous Hall resistance. The current pulse has a rise time of ~70 ps and a fall time of ~80 ps, and its magnitude is determined by measuring the transmitted signal. The switching probability under current pulses is obtained from the following procedure: we applied a negative reset DC-current of 1.5 mA to initialize the magnetization to 'up' state followed by a positive current pulse for SOT switching. A few seconds after each pulsed-current, the anomalous Hall resistance is measured using a low DC current of +70 μA to sense the magnetization state. Individual current pulse injections with a fixed amplitude $J$ and duration $t$ were repeated 20 times to determine the switching probability which is defined as $P_{sw}$ = (number of 'down' states) / 20.

We studied total 9 devices with varying the dot diameter and ferromagnet thickness. Every device showed the backward switching with quantitative difference in the switching phase diagram that is attributed to the deviation in effective anisotropy and depinning sites.

**Intrinsic switching current density from the macrospin switching model** The intrinsic switching current density from the macrospin-like rotation switching model is calculated[46] using

$$J_{c0} = \frac{2e}{\hbar} \frac{M_s t_F}{\theta_{SH}} \left( \sqrt{\frac{H_{k,eff}^2}{32} \left[ 8 + 20\left(\frac{H_x}{H_{k,eff}}\right)^2 - \left(\frac{H_x}{H_{k,eff}}\right)^4 - \left(\frac{H_x}{H_{k,eff}}\right)\left(8 + \left(\frac{H_x}{H_{k,eff}}\right)^2\right)^{3/2}\right]} \right). \quad (1)$$

This rotational switching model gives $J_{c0}$ of $148 \times 10^6$ A cm$^{-2}$ using parameters of the saturation magnetization $M_s$ = 670 emu/cm$^3$ (from VSM measurement), the perpendicular anisotropy $H_{k,eff}$ = 3000 Oe (from VSM measurement), the ferromagnetic layer thickness $t_F$ = 0.9 nm, spin Hall angle $\theta_H$ = 0.09, and in-plane assist field $H_x$ = 1191 Oe.

**1D model calculation** Micromagnetics simulations are carried out by numerically solving the Landau–Lifshitz–Gilbert equation[47] including the damping-like and field-like component of spin-orbit torque:

$$\frac{d\hat{\mathbf{m}}}{dt} = -\gamma\mu_0\hat{\mathbf{m}} \times \hat{\mathbf{H}}_{eff} + \alpha\hat{\mathbf{m}} \times \frac{d\hat{\mathbf{m}}}{dt} - \gamma\hat{\mathbf{m}} \times (-\tau_{DLT}\hat{\mathbf{m}} \times \hat{\mathbf{y}}) - \gamma\hat{\mathbf{m}} \times (-\tau_{FLT}\hat{\mathbf{y}}), \quad (2)$$

where $\tau_{DLT} = c_{DLT}(\hbar J)/(2eM_s d)$ and $\tau_{FLT} = c_{FLT}(\hbar J)/(2eM_s d)$. The equivalent field for each torque terms are defined as $H_{DLT} = -\tau_{DLT}(\hat{\mathbf{m}} \times \hat{\mathbf{y}})$ and $H_{FLT} = -\tau_{FLT}\hat{\mathbf{y}}$. $\hat{\mathbf{H}}_{eff}$ is the effective field including the magnetostatic field, anisotropy field, exchange field, Dzyaloshinskii-Moriya interaction field, and external field. Following parameters are used: the saturation magnetization $M_s$ = 670 emu cm$^{-3}$ (from VSM measurement, see Supplementary Note 9), the perpendicular anisotropy $K$ = 3.83 × 10$^6$ erg cm$^{-3}$ (from VSM measurement), the Dzyaloshinskii-Moriya interaction constant $D$ = 0.05 erg cm$^{-2}$, the exchange stiffness constant $A_{ex}$ = 2.0 × 10$^{-6}$ erg cm$^{-1}$, the damping $\alpha$ = 0.035 (from FMR measurement, see Supplementary Note 10), the DLT efficiency $c_{DLT}$ = –0.09, and the FLT efficiency $c_{FLT}$ = +0.29 (from the harmonic measurement, see Supplementary Note 8). The corresponding FLT/DLT ratio is −3.2. For the current pulse, both rise and fall times are 100 ps. In our sign convention, a negative DLT efficiency ($c_{DLT} < 0$) induces an 'up'-to-'down' switching for $J > 0$ and $H < 0$ ($\theta_H = 180°$). The considered geometry has dimension of 220 nm × 80 nm × 2.5 nm and the unit cell size of 2 nm × 80 nm × 2.5 nm. The initial magnetization direction of majority part of the considered geometry is 'up' (along +z-direction) with reversed 'down' domain formed at one edge. The DMI effective field ($H_{DMI}$) is estimated using the following relation[30, 42]: $H_{DMI} = D/\Delta M_s$. Here, $D = +0.05$ erg cm$^{-2}$ is the reported DMI constant value in the Ta/CoFeB/MgO structure[38, 42, 48], and $\Delta = \sqrt{A_{ex}/K}$ is the DW width.

13. Pai, C.-F. *et al.* Spin transfer torque devices utilizing the giant spin Hall effect of tungsten. *Appl. Phys. Lett.* **101,** 122404 (2012).
14. Emori, S., Bauer, U., Ahn, S.-M., Martinez, E. & Beach, G. S. D. Current-driven dynamics of chiral ferromagnetic domain walls. *Nat. Mater.* **12,** 611-616 (2013).
15. Yamanouchi, M. *et al.* Three terminal magnetic tunnel junction utilizing the spin Hall effect of iridium-doped copper. *Appl. Phys. Lett.* **102,** 212408 (2013).
16. Lee, S.-W. & Lee, K.-J. Effect of Wavy Spin-Transfer Torque on Ultrafast Unipolar Switching of a Nanomagnet With Out-of-Plane Polarizer. *IEEE Trans. Magn.* **47,** 3872-3875 (2011).
17. Kent, A. D., Özyilmaz, B. & Barco, E. d. Spin-transfer-induced precessional magnetization reversal. *Appl. Phys. Lett.* **84,** 3897 (2004).
18. Papusoi, C. *et al.* 100 ps precessional spin-transfer switching of a planar magnetic random access memory cell with perpendicular spin polarizer. *Appl. Phys. Lett.* **95,** 072506 (2009).
19. Kanai, S. *et al.* Magnetization switching in a CoFeB/MgO magnetic tunnel junction by combining spin-transfer torque and electric field-effect. *Appl. Phys. Lett.* **104,** 212406 (2014).
20. Park, J., Rowlands, G. E., Lee, O. J., Ralph, D. C. & Buhrman, R. A. Macrospin modeling of sub-ns pulse switching of perpendicularly magnetized free layer via spin-orbit torques for cryogenic memory applications. *Appl. Phys. Lett.* **105,** 102404 (2014).
21. Legrand, W., Ramaswamy, R., Mishra, R. & Yang, H. Coherent Subnanosecond Switching of Perpendicular Magnetization by the Fieldlike Spin-Orbit Torque without an External Magnetic Field. *Phys. Rev. Appl.* **3,** 064012 (2015).
22. Kanai, S. *et al.* In-plane magnetic field dependence of electric field-induced magnetization switching. *Appl. Phys. Lett.* **103,** 072408 (2013).
23. Finocchio, G., Carpentieri, M., Martinez, E. & Azzerboni, B. Switching of a single ferromagnetic layer driven by spin Hall effect. *Appl. Phys. Lett.* **102,** 212410 (2013).
24. Garello, K. *et al.* Ultrafast magnetization switching by spin-orbit torques. *Appl. Phys. Lett.* **105,** 212402 (2014).
25. Fukami, S., Anekawa, T., Zhang, C. & Ohno, H. A spin–orbit torque switching scheme with collinear magnetic easy axis and current configuration. *Nat. Nanotech.* **11,** 621-625 (2016).
26. Fan, X. *et al.* Observation of the nonlocal spin-orbital effective field. *Nat. Commun.* **4,** 1799 (2013).
27. Garello, K. *et al.* Symmetry and magnitude of spin-orbit torques in ferromagnetic heterostructures. *Nat. Nanotech.* **8,** 587-593 (2013).
28. Skinner, T. D. *et al.* Spin-orbit torque opposing the Oersted torque in ultrathin Co/Pt bilayers. *Appl. Phys. Lett.* **104,** 062401 (2014).
29. Qiu, X. *et al.* Spin–orbit-torque engineering via oxygen manipulation. *Nat. Nanotech.* **10,** 333-338 (2015).
30. Lee, O. J. *et al.* Central role of domain wall depinning for perpendicular magnetization switching driven by spin torque from the spin Hall effect. *Phys. Rev. B* **89,** 024418 (2014).
31. Zhang, C., Fukami, S., Sato, H., Matsukura, F. & Ohno, H. Spin-orbit torque induced magnetization switching in nano-scale Ta/CoFeB/MgO. *Appl. Phys. Lett.* **107,** 012401 (2015).16

**Acknowledgments**

This research was supported by the National Research Foundation (NRF), Prime Minister's Office, Singapore, under its Competitive Research Programme (CRP award no. NRFCRP12-2013-01). J.M.L. thanks S.-W. Lee and K.-J. Lee for useful discussions and J. Yu for helping graphic works.

**Author contributions**
J.M.L., J.H.K., and H.Y. initiated the project. J.M.L., X.Q. and R.M. deposited and characterized films. J.M.L., J.H.K., and J.S. fabricated devices. J.M.L., J.H.K. and J.B.Y. performed measurements. J.M.L. did micromagnetic simulations. S.S. measured the FMR. K.C. conducted the calculations based on the finite element method. All authors discussed the results. J.M.L., R.R, and H.Y. wrote the manuscript. H.Y. supervised the project.

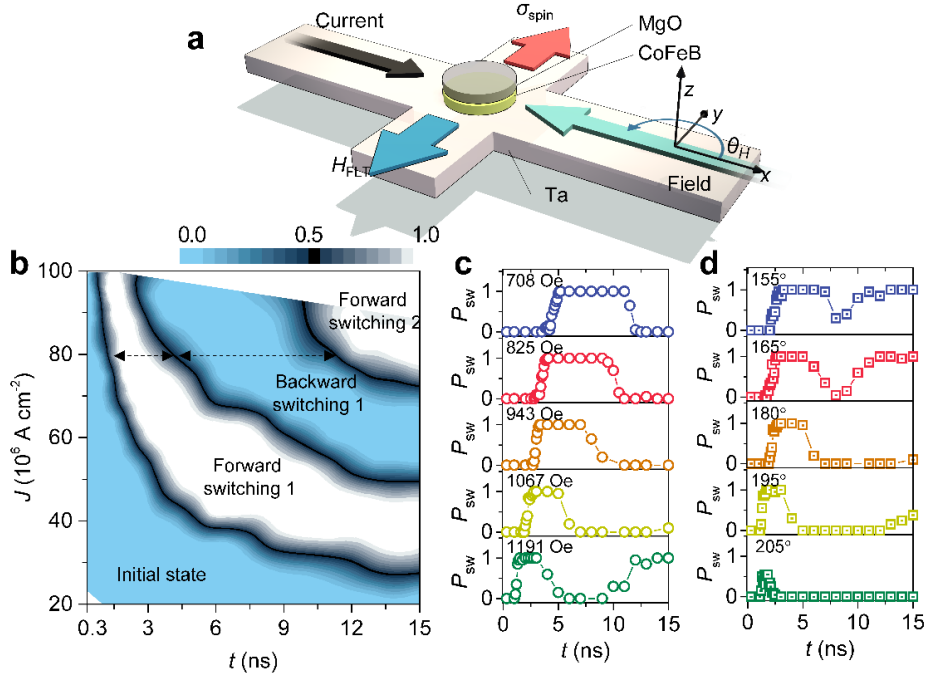

**Figure 1 | Spin-orbit torque switching probability diagram** (**a**) Schematic diagram of spin-orbit torque device. CoFeB circular dot with perpendicular magnetic anisotropy is on top of Ta Hall-cross. DC and current pulses are injected along the Ta channel (*x*-axis) and assist-field is applied in-plane (*xy* plane). Arrows in blue and red denote the direction of spin polarization and field-like torque equivalent field, respectively. (**b**) Switching probability diagram with respect to the current density (*J*) and pulse duration (*t*), with assist field (*H*) of 1191 Oe. The scale bar indicates the contour level of switching probability (×100 %). The device has an initial 'up' magnetization configuration. The forward and backward switching indicates 'up' to 'down' and 'down' to 'up' switching, respectively. The dashed arrows indicate the periods of forward switching and backward switching for an applied *J* of $79.4 \times 10^6$ A cm$^{-2}$, respectively. (**c**) The switching probability (×100 %) as a function of *t* with varying *H*. *J* and $\theta_H$ are fixed to $79.4 \times 10^6$ A cm$^{-2}$ and 180°, respectively. (**d**) Switching probability (×100 %) as a function of *t* with varying in-plane angle of assist-field ($\theta_H$). *J* and *H* are fixed to $79.4 \times 10^6$ A cm$^{-2}$ and 1067 Oe, respectively.

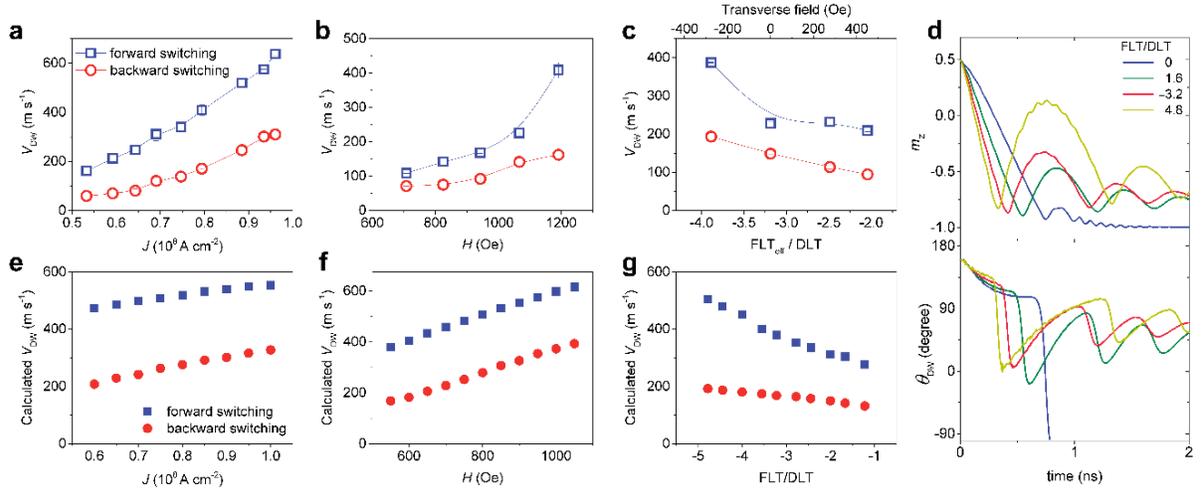

**Figure 2 | Spin-orbit torque driven domain wall dynamics** (**a**,**b**,**c**) Estimated domain wall velocity ($V_{DW}$) during the forward switching and backward switching with respect to (**a**) current density $J$ with a fixed assist field $H = 1191$ Oe, (**b**) $H$ with a fixed $J = 79.4 \times 10^6$ A cm$^{-2}$, and (**c**) FLT$_{eff}$/DLT ratio with a fixed $J = 79.4 \times 10^6$ A cm$^{-2}$ and $H = 1067$ Oe. Error bars indicate the standard deviation from several measurements. (**d**) Temporal evolutions of out-of-plane magnetization $m_z$ (top panel) and the domain wall angle $\theta_{DW}$ (bottom panel) under spin-orbit torque from 1D model. (**e**,**f**,**g**) Calculated domain wall velocity during the forward switching and backward switching with respect to (**e**) current pulse density $J$, (**f**) in-plane assist field $H$, and (**g**) FLT/DLT ratio.

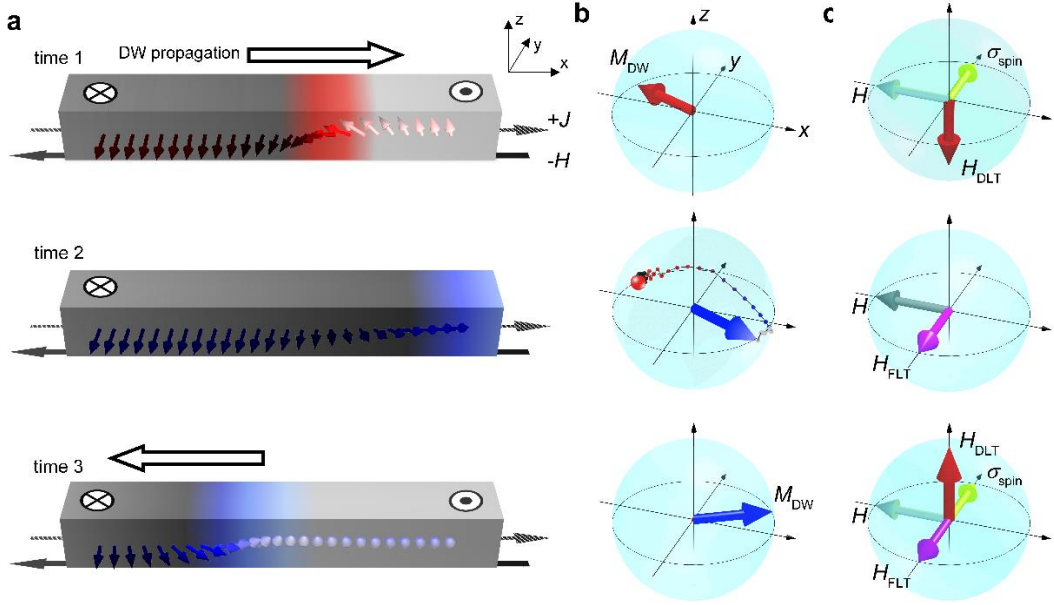

**Figure 3 | Schematic illustrations of spin-orbit torque switching via domain expansion** (**a**) Domain wall configuration, (**b**) internal magnetization at the center of domain wall ($M_{DW}$), and (**c**) the dominant torques acting on the domain wall during the SOT switching process. Time 1: applying current results in the initial 'up' to 'down' forward switching as $H_{DLT}$ acting on domain wall ($\mathbf{M_{DW}} \cdot \hat{\mathbf{x}} < 0$) expands the 'down' domain. Time 2: Domain wall annihilates at the structure edge followed by nucleation of a domain wall with inverted chirality ($\mathbf{M_{DW}} \cdot \hat{\mathbf{x}} > 0$) over time. Time 3: the 'down' to 'up' backward switching follows the forward switching in time as $H_{DLT}$ acting on domain wall ($\mathbf{M_{DW}} \cdot \hat{\mathbf{x}} > 0$) expands the 'up' domain. Current and assist-field are along the +x and –x-direction, respectively, and time 1 < time 2 < time 3. The light and dark gray colors in (**a**) indicate the out-of-plane component of magnetization ($\pm M_z$), while the blue and red colors indicate the longitudinal component of magnetization ($\pm M_x$).

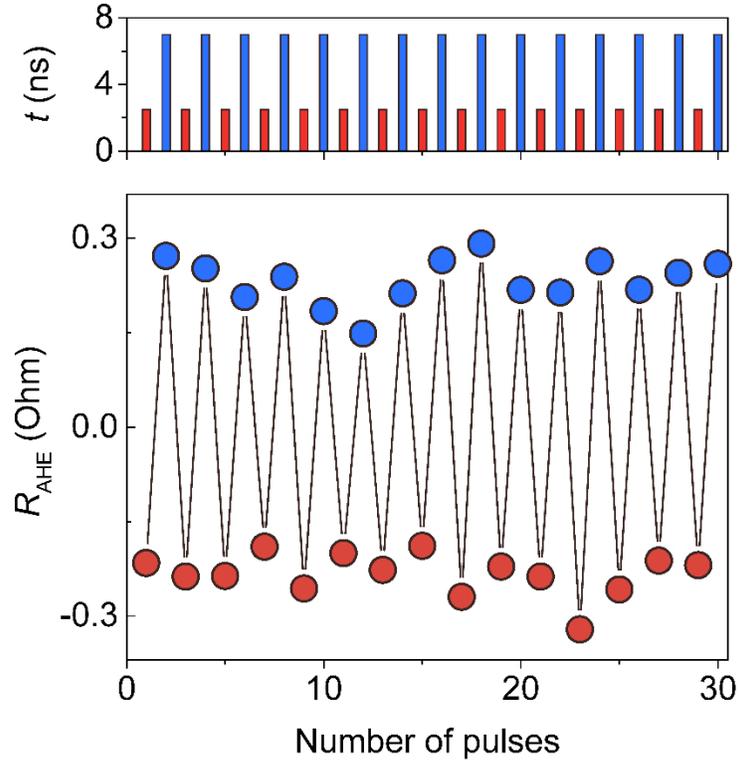

**Figure 4 | Unipolar spin-orbit torque switching** Unipolar switching of CoFeB dot with perpendicular magnetic anisotropy by applying a series of positive current pulses with alternating duration of 2.5 ns and 7.5 ns. The positive and negative values of anomalous Hall resistance ($R_{AHE}$) indicate the magnetization with 'up' and 'down' configuration, respectively. A constant current pulse density of $79.4 \times 10^6$ A cm$^{-2}$ (along the +$x$-direction) and magnetic field of 1067 Oe (along the –$x$-direction) were applied. The magnetization was initialized after each current pulse injection.